\documentclass{ws-procs975x65}
\usepackage{epstopdf}
\begin{document}

\title{Exotic phases and quantum phase
transitions:\\ model systems and experiments}

\author{Subir Sachdev}
  
\address{Department of Physics, Harvard University, Cambridge MA 02138}

\begin{abstract}
I survey theoretical advances in our understanding of the quantum phases and phase transitions
of Mott insulators, and of allied conducting systems obtained by doping charge carriers. 
A number of new experimental examples of Mott insulators have appeared in recent years,
and I critically compare their observed properties with the theoretical expectations.
\end{abstract}

\bodymatter
\begin{center}
{\tt Rapporteur talk at the 24th Solvay Conference on Physics,\\ {\em Quantum Theory of 
Condensed Matter\/}, Brussels, Oct 11-13, 2008}
\end{center}
\section{Introduction}
\label{sec:intro}

The band theory of electrons predicts that any crystal with an odd number of electrons per
unit cell must be a metal. However, strong electron-electron interactions can invalidate this conclusion,
and such crystals can also be insulators, known as Mott insulators.
I will use this term here more broadly: often the Mott insulator has a secondary instability
to spin or charge ordering which increases the size of the unit cell, so that the ultimate ground state of
the insulator does have an even number of electrons per unit cell (many examples of such instabilities will
be discussed below).  I will continue to refer to such insulators as Mott insulators because electron-electron
interactions are crucial to understanding their broken symmetries and excitation spectrum. In contrast, describing the ordering
by using the instabilities of a metallic state with an odd number of electrons per unit cell leads to a rather poor
understanding of the insulator and of the energy scales characterizing its excitations.

A canonical model used to study Mott insulators is the single-band Hubbard model
\begin{equation}
H_U = \sum_{i,j} \left( - t_{ij} - \mu \delta_{ij} \right) 
c^{\dagger}_{i \alpha} c_{j \alpha} + U \sum_i n_{i \uparrow} n_{i \downarrow}
\label{HU}
\end{equation}
where $c_{i \alpha}$ annihilates an electron with spin $\alpha = \uparrow,\downarrow$
on the sites, $i$, of a regular lattice, and $n_{i \alpha} = c^{\dagger}_{i \alpha} c_{i \alpha}$ is the number
operators for these electrons. For small $U$, the ground state of $H_U$ is a metal (on most lattices) which can be
described in the traditional framework of band and Fermi liquid theory.
For strong repulsion between the electrons with $U /|t_{ij}| \gg 1$, and with the chemical potential
$\mu$ adjusted so that there is one electron per unit cell,
charge fluctuations are strongly suppressed on each site, and the ground state is a Mott insulator.
The low energy excitations of the Mott insulator are described by an effective Hamiltonian
which is projected onto the subspace of states with exactly one electron per site. These states are described
by the spin orientation of each electron, and the effective Hamiltonian is a Heisenberg quantum spin model
\begin{equation}
H_J = \sum_{i<j} J_{ij} {\bf S}_i \cdot {\bf S}_j + \ldots
\label{HJ}
\end{equation}
where $J_{ij} = 4 t_{ij}^2/U$ is the antiferromagnetic exchange interaction, ${\bf S}_i$ is the spin $S=1/2$ operator
on site $i$, and the ellipses refer to multiple spin-exchange terms which are generated at higher orders 
in the expansion in $t_{ij}/U$. One of the purposes of this article is to survey
theoretical advances in understanding the ground states of $H_J$ on a variety of lattices
in two spatial dimensions. A number of experimental realizations (some newly discovered)
will also be surveyed and critically compared with theory.

More broadly, the study of models like $H_U$ and $H_J$ will lead us to a number of exotic phases, both insulating
and conducting,
which require modern concepts from gauge theory and `topological' order for their complete characterization.
Our unifying strategy here will be to access these exotic states across a quantum phase transition from a conventional
state. We will begin by characterizing the `order' in a conventional state, and then turn up the strength of 
quantum fluctuations leading to a quantum `disordering' transition to an exotic state. This approach will lead to 4 broad
classes of exotic states, discussed in the sections below:\\
({\em i\/}) \underline{\em Quantum fluctuating antiferromagnetism}. We begin with an insulator with antiferromagnetic long-range order, well described by $H_J$.
Quantum fluctuations of the antiferromagnetism lead to states with full SU(2) spin rotation symmetry,
and an energy gap to spin excitations. 
In Section~\ref{sec:lgw} we consider a simple, and now well-understood model:
the coupled dimer antiferromagnet. In this case, well-developed methods from the theory of finite
temperature phase transitions can be extended to successfully describe its ground states
and quantum phase transition.  In Section~\ref{sec:tri}, we will introduce a recent experimental
example of a triangular lattice antiferromagnet, and develop a theory for the non-magnetic insulating 
states in which the quantum interference effects play a more fundamental role, and new theoretical
ideas are required.\\
({\em ii\/}) \underline{\em Neutral fermions across the Mott transition}. 
We begin with the Fermi liquid state of $H_U$, characterized by a Fermi surface (in some cases, Fermi points)
of charge $\pm e$, spin $S=1/2$ quasiparticles. Now we postulate a continuous Mott transition to an insulator
in which the spin and charge of the quasiparticles separate, and a `ghost' Fermi surface survives in the insulator,
with the Fermi surface excitations carrying $S=1/2$ spin, but no charge; these are fermionic spinons. 
The current status of such exotic states
will be reviewed in Section~\ref{sec:mott}.\\
({\em iii\/}) \underline{\em Breakdown of Kondo screening}.
As discussed in Section~\ref{sec:ffl}, the heavy fermion state of rare-earth intermetallics is described by 
the Kondo-Heisenberg model describing the exchange coupling of local moments to itinerant conducting electrons.
The Kondo effect tightly entangles the local spins and the itinerant electrons, leading to a `large Fermi surface'
state, which encloses a volume determined by the total electron density, including both the local and itinerant electrons.
For sufficiently strong exchange between the spins, the Kondo screening can break down, and the local moments
and itinerant electrons disentangle, leading to a `fractionalized Fermi liquid'. 
In the simplest models, the itinerant electrons form a small, metallic
Fermi surface of conventional electronic quasiparticles, while the local moments form a spinon Fermi surface.\\
({\em iv\/}) \underline{\em Quantum fluctuating metallic spin density waves}.
We begin with a metallic Fermi liquid state, in the presence of spin density wave order.
This order will generally break up the Fermi surface into electron and hole pockets. Section~\ref{sec:acl}
will describe a quantum transition in which the spin density wave order becomes short range,
but ghost Fermi pockets survive in the resulting `algebraic charge liquid'. In the latter state, the Fermi surface excitations
carry charge $\pm e$ but no spin. Such a state has been used recently to develop a theory of the enigmatic
underdoped region of the cuprates. 

The concluding Section~\ref{sec:exp} will survey recent experiments on Mott insulators on
a number of frustrated lattices, and compare observations with numerical studies and the theoretical
proposals.

\section{Coupled dimer antiferromagnet} 
\label{sec:lgw}

\begin{figure}
\begin{center}
 \includegraphics[width=4.5in]{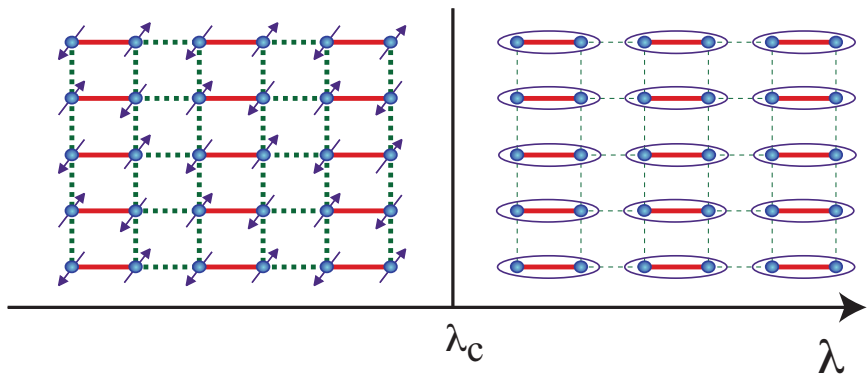}
 \caption{The coupled dimer antiferromagnet. The full red lines represent an exchange interaction $J$,
 while the dashed green lines have exchange $J/\lambda$. The ellispes represent a singlet valence
 bond of spins $(|\uparrow \downarrow \rangle - | \downarrow \uparrow \rangle )/\sqrt{2}$.}
\label{fig:dimer}
\end{center}
\end{figure}

This model is illustrated in Fig.~\ref{fig:dimer}. 
The $S=1/2$ spins reside on the sites of a square lattice, and have nearest neighbor exchange equal
to either $J$ or $J/\lambda$. Here $\lambda \geq 1$ is a tuning parameter which induces a quantum phase
transition in the ground state of this model. 

At $\lambda = 1$, the model has full square lattice symmetry,
and this case is known to have a N\'eel ground state which breaks spin rotation symmetry. This state has a
checkerboard polarization of the spins, just as found in the classical ground state, and as illustrated 
on the left side of Fig.~\ref{fig:dimer}. It can be characterized by a vector order parameter ${\bm \varphi}$
which measures the staggered spin polarization
\begin{equation}
{\bm \varphi} = \eta_i {\bf S}_i
\end{equation}
where $\eta_i=\pm 1$ on the two sublattices of the square lattice. In the N\'eel state we have $\langle {\bm \varphi} \rangle \neq  0$,
and we expect that the low energy excitations can be described by long wavelength fluctuations of a field ${\bm \varphi} (x, \tau)$ over
space, $x$, and imaginary time $\tau$.

On the other hand, for $\lambda \gg 1$ it is evident from Fig.~\ref{fig:dimer} that the ground state preserves 
all symmetries of the Hamiltonian: it has total spin $S=0$ and can be considered to be a product of nearest
neighbor singlet valence bonds on the $J$ links. It is clear that this state cannot be smoothly connected
to the N\'eel state, and so there must at least one quantum phase transition as a function $\lambda$. 

Extensive quantum Monte Carlo simulations \cite{troyer,matsu,janke} 
on this model have shown there is a direct phase
transition between these states at a critical $\lambda_c$, as in Fig.~\ref{fig:dimer}. These simulations have
no sign problem, and so it has been possible to obtain extremely precise results. The value of $\lambda_c$
is known accurately, as are the critical exponents characterizing a second-order quantum phase
transition. These critical exponents are in excellent agreement with the simplest proposal for the critical
field theory, \cite{janke} which can be obtained via conventional Landau-Ginzburg arguments. Given the vector
order parameter ${\bm \varphi}$, we write down the action in $d$ spatial and one time dimension,
\begin{equation}
\mathcal{S}_{LG} = \int d^d r d\tau \left[ \frac{1}{2} \left[ (\partial_\tau {\bm \varphi} )^2  + c^2 ( \nabla {\bm \varphi} )^2 + s {\bm \varphi}^2 \right]
+ \frac{u}{4} \left[ {\bm \varphi}^2 \right]^2 \right], \label{slg}
\end{equation}
as the simplest action expanded in gradients and powers of ${\bm \varphi}$ which is consistent will all
the symmetries of the lattice antiferromagnet.
The transition is now tuned by varying $s \sim (\lambda - \lambda_c)$. Notice that this model 
is identical to the Landau-Ginzburg theory for the thermal phase transition in a $d+1$ dimensional ferromagnet,
because time appears as just another dimension. As an example of the agreement: the critical exponent of the correlation
length, $\nu$, has the same value, $\nu = 0.711 \ldots$, to three significant digits in a quantum Monte Carlo study of the coupled
dimer antiferromagnet,\cite{janke} and in a 5-loop analysis \cite{vicari} of the renormalization group fixed point of $\mathcal{S}_{LG}$
in $d=2$. 
Similar excellent agreement is obtained for the double-layer antiferromagnet \cite{sandsca,matsushita}
and the coupled-plaquette antiferromagnet.\cite{afa}

\begin{figure}
\begin{center}
 \includegraphics[width=3.5in]{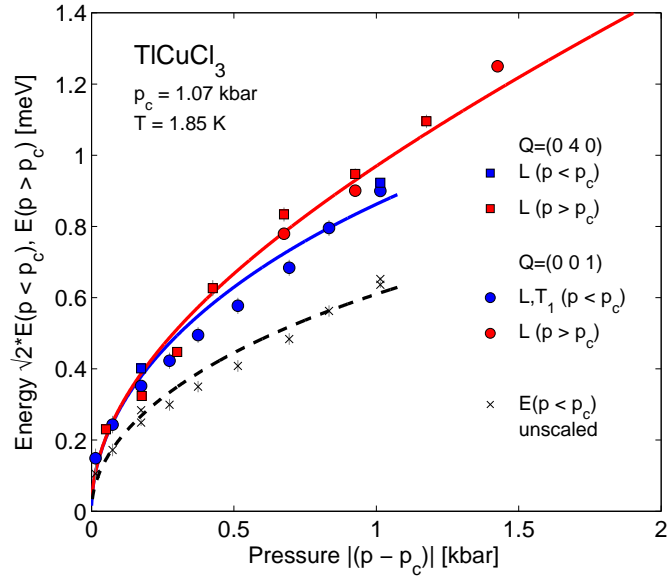}
 \caption{Energies of the gapped collective modes across the pressure ($p$) tuned quantum phase transition in
 TlCuCl$_3$ observed by Ruegg {\em et al.}\cite{ruegg}. We test the description by the action $\mathcal{S}_{LG}$
 in Eq.~(\ref{slg}) with $s \propto (p_c - p)$ by comparing $\sqrt{2}$ times the energy gap for $p<p_c$ with the
 energy of the longitudinal mode for $p>p_c$. The lines are the fits to a $\sqrt{|p-p_c|}$ dependence, testing the 1/2
 exponent.}
\label{fig:ruegg}
\end{center}
\end{figure}

In experiments, the best studied realization of the coupled-dimer antiferromagnet is TlCuCl$_3$. In this crystal, the dimers are coupled
in all three spatial dimensions, and the transition from the dimerized state to the N\'eel state can be induced by application of pressure.
Neutron scattering experiments by Ruegg and collaborators \cite{ruegg} have 
clearly observed the transformation in the excitation spectrum across the transition,
as is described by a simple fluctuations analysis about the mean field saddle point of $\mathcal{S}_{LG}$. In the dimerized phase
($s>0$), a triplet of gapped excitations is observed, corresponding to the three normal modes of ${\bm \varphi}$ oscillating
about ${\bm \varphi} = 0$; as expected, this triplet gap vanishes upon approaching the quantum critical point. In a mean field analysis,
valid for $d \geq 3$,
the field theory in Eq.~(\ref{slg}) has a triplet gap of $\sqrt{s}$.
In the N\'eel phase, the neutron scattering detects 2 gapless spin waves, and one gapped longitudinal
mode \cite{normand} (the gap to this longitudinal mode vanishes at the quantum critical point), as is expected from fluctuations
in the inverted `Mexican hat' potential of $\mathcal{S}_{LG}$ for $s<0$. The longitudinal mode has a mean-field
energy gap of $\sqrt{2 |s|}$.
These mean field predictions for the energy of the gapped modes on the two sides of the transition are tested in Fig.~\ref{fig:ruegg}:
the observations are in good agreement with the 1/2 exponent and the 
predicted \cite{sslg} $\sqrt{2}$ ratio, providing a non-trival experimental
test of the $\mathcal{S}_{LG}$ field theory. 

\section{Quantum ``disordering'' magnetic order: spinons and visons}
\label{sec:tri}

\begin{figure}
\begin{center}
 \includegraphics[width=2.2in]{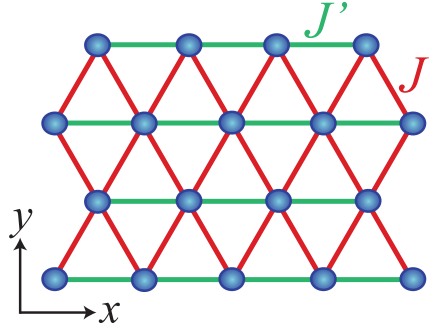}
 \caption{The antiferromagnet on the distorted triangular lattice with exchange couplings
 $J$ and $J'$.}
\label{fig:tri}
\end{center}
\end{figure}

Now consider the triangular lattice antiferromagnet illustrated in Fig.~\ref{fig:tri},
with nearest neighbor exchange constants $J$ and $J'$. 
A fundamental difference from Section~\ref{sec:lgw} is that now there is only one
site per unit cell. Consequently, it is not as simple to write down a simple quantum ``disordered'
state, such as the large $\lambda$ dimerized state in Fig.~\ref{fig:dimer}; any single pairing of the electrons
into singlet bonds must break the lattice translations symmetry of the Hamiltonian, unlike the situation
for the coupled dimer antiferromagnet. We will see that this difficulty leads to a great deal of complexity,
and a rich class of field theories which can describe the quantum spin fluctuations.

A seemingly simple and well-posed
problem is to describe the ground state of $H_J$ for this lattice as a function of $J' /J$. 
However, the answer to this question is not known with anywhere close to the reliability
of the model in Section~\ref{sec:lgw}. The main reason is that the sign problem prevents
large scale Monte Carlo simulations, and we have to rely on series expansions, \cite{wms} exact diagonalizations
on relatively small systems, \cite{misguich1,misguich2} 
or the recently developed variational approach based upon PEPS states. \cite{cirac}
Below, we will review theoretical proposals based upon an approach which begins from the 
ground state of the classical antiferromagnet, and attempts to quantum ``disorder'' it by a systematic
analysis of the quantum fluctuations in its vicinity.\cite{rsl,rstl,sst,css}

\begin{figure}
\begin{center}
 \includegraphics[width=4.5in]{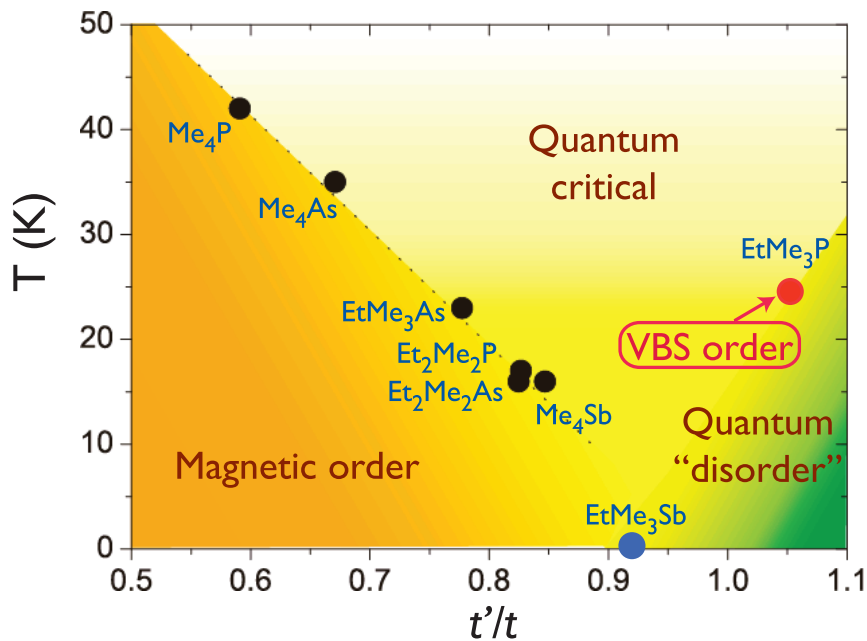}
 \caption{Phase diagram of X[Pd(dmit)$_2$]$_2$
 from Shimizu {\em et al.}~\cite{kato6}. 
Each point is identified with the cation X. The values of the ratio of electron hopping, $t'/t$ were
obtained from quantum chemistry computatons; the exchange interactions $J'/J \approx (t'/t)^2$. 
The black points on the left represent compounds with antiferromagnetic order, and they
are placed at the magnetic ordering temperature. The red point, EtMe$_3$P, is in antiferromagnet with a spin gap
which acquires valence bond solid (VBS) order at the indicated temperature. Finally, the blue point, EtMe$_3$Sb,
is a compound for which no order has been discovered so far.}
\label{fig:kato}
\end{center}
\end{figure}

Experimental motivation for the antiferromagnet illustrated in Fig.~\ref{fig:tri} comes
from a remarkable series of experiments by the group of Reizo Kato \cite{kato1,kato2,kato3,kato4,kato5,kato6,kato7}
on the organic
Mott insulators X[Pd(dmit)$_2$]$_2$. These insulators crystallize in a layered structure,
with each layer realizing a copy of the triangular lattice in Fig.~\ref{fig:tri}.
Each site of this
lattice has a pair of Pd(dmit)$_2$ molecules
carrying charge $-e$ and spin $S=1/2$, which then interact antiferromagnetically with each
other with exchange constants $J$ and $J'$. 
The ingredient
X intercalates between the triangular layers, and can range over a variety of
monovalent cations. The choice of X yields a powerful experimental tuning knob, 
because different X correspond to different values of 
$J'/J$. The current status of experiments on these compounds is summarized in the phase
diagram in Fig.~\ref{fig:kato}.
For small $J'/J$ antiferromagnetic order is observed in NMR experiments. Indicated in Fig.~\ref{fig:kato} is the
magnetic ordering temperature: the long-range magnetic order is a consequence of the weak inter-layer
coupling. For $J'=0$, the lattice in Fig.~\ref{fig:tri} is equivalent to the $\lambda=1$ square lattice in Fig.~\ref{fig:dimer}, 
and so the antiferromagnetic order is expected to have the two-sublattice N\'eel structure shown in the left
panel of Fig.~\ref{fig:dimer}. It is evident from Fig.~\ref{fig:kato} that the strength of the antiferromagnetic
order decreases with increasing $J' /J$ until ultimately yielding a quantum `disordered' state. The nature
of the latter state and of the quantum critical point to the magnetically ordered state are among the key issues
we wish to address here.

\begin{figure}
\begin{center}
 \includegraphics[width=2.5in]{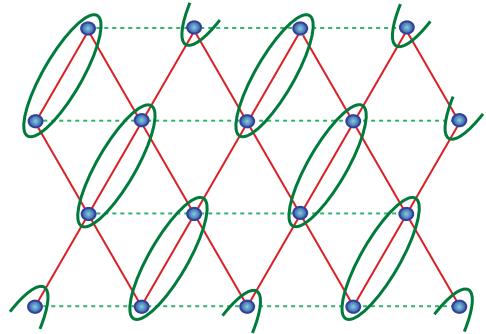}
 \caption{Schematic of the valence bond solid (VBS) found in  X[Pd(dmit)$_2$]$_2$ 
 for X=EtMe$_3$P. The ellipses represent singlet bonds, as in Fig.~\ref{fig:dimer}. }
\label{fig:vbs}
\end{center}
\end{figure}

Experiments on the  X[Pd(dmit)$_2$]$_2$ Mott insulators indicate a possible structure of the
quantum `disordered' state. As indicated in Fig.~\ref{fig:kato}, the compound with X=EtMe$_3$P
has valence bond solid (VBS) order in a state with a spin gap.\cite{kato4,kato5} This is a state with an gap to all non-zero
spin excitations of $\approx$ 40~K, as measured by an exponential suppression of the spin susceptibility.
Below a temperature $\approx$ 26~K there is a doubling of the unit cell, consistent with the ordering
of the singlet valence bonds as indicated in Fig.~\ref{fig:vbs}. 
Note that the wavefunction of this state appears
similar to the coupled dimer state in the right panel of Fig.~\ref{fig:dimer}. However, the crucial difference
is that the valence bond ordering pattern is not imposed by the Hamiltonian but is due to a spontaneously broken
symmetry. There are 4 equivalent valence bond ordering patterns with an energy identical to the state in 
Fig.~\ref{fig:vbs}, obtained by operating on it by translational and rotational symmetries of the lattice.
Theoretically, such a state was predicted \cite{rsl} to exist proximate to the N\'eel state, with the symmetry
breaking arising as a consequence of quantum Berry phases which are {\em not\/} present in the theory
of the coupled-dimer antiferromagnet in Eq.~(\ref{slg}).
We note in passing that recent scanning tunneling experiments on the underdoped cuprates have also displayed evidence
for VBS-like correlations.\cite{kohsaka1,kohsaka2}

Also of interest in Fig.~\ref{fig:kato} is the compound with X=EtMe$_3$Sb which has no apparent ordering \cite{kato7}
and so is in a ``spin liquid'' state. Its placement in the phase diagram in Fig.~\ref{fig:kato} indicates
that an appropriate description might between in terms of a quantum critical point between the ordered phases.
Its properties appear similar to another well-studied spin liquid compound, $\kappa$-(ET)$_2$Cu$_2$(CN)$_3$,
which will be discussed in Section~\ref{sec:triangle}.

The subsections below will review the extensions needed to extend $\mathcal{S}_{LG}$ in 
Eq.~(\ref{slg}) to be a  complete theory of two-dimensional quantum antiferromagnets with a single
$S=1/2$ spin per unit cell. A useful way of developing this extension is to postulate a spin liquid state
in the form originally envisaged by Pauling\cite{pauling} and Fazekas and Anderson:\cite{fazekas} a state which is a superposition of a large
number of singlet bond pairings of the electrons (of which the pairing in Fig.~\ref{fig:vbs} in just one)
in a manner which preserves all the symmetries of the lattice. Such a state has two primary classes of excitations,
spinons and visons, whose properties are reviewed below. As we will see, a rich variety of ordered phases
and critical points are obtained when we allow one or more of these excitations to condense.

\subsection{Spinons}
\label{sec:spinons}

Returning to our picture of quantum `disordering' the N\'eel state, the key step \cite{arovas} is to 
replace our vector order parameter ${\bm \varphi}$ by a two-component bosonic spinor $z_\alpha$ ($\alpha = \uparrow, \downarrow$)
\begin{equation}
{\bm \varphi} = z_\alpha^\ast {\bm \sigma}_{\alpha \beta} z_\beta ,
\label{neel}
\end{equation}
where ${\bm \sigma}$ are the Pauli matrices. We are clearly free to describe quantum fluctuations of the N\'eel order in
terms of the complex doublet $z_\alpha$ rather than the real vector ${\bm \varphi}$. However, the new description is redundant:
a spacetime-dependent U(1) gauge transformation
\begin{equation}
z_\alpha \rightarrow e^{i \phi} z_\alpha
\end{equation}
leaves the observable ${\bm \varphi}$ invariant, and so should lead to a physically equivalent state. Any local effective action
for the $z_\alpha$ must be invariant under this gauge transformation, and so we are led to introduce an `emergent' U(1) gauge
field $a_\mu$ to facilitate local gradient terms in such an action. This proliferation of degrees of freedom from the previous economical
description in terms of ${\bm \varphi}$ might seem cumbersone, but it ultimately allows for the most efficient description of all
the excitations, and their Berry phases.

\begin{figure}
\begin{center}
 \includegraphics[width=2.5in]{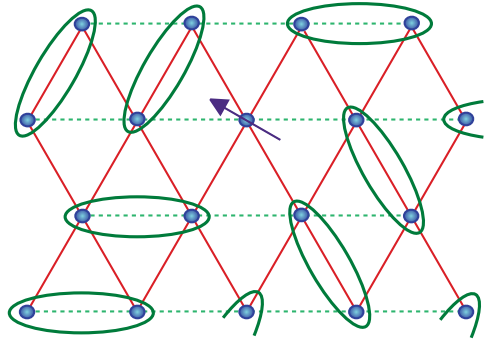}
 \caption{Schematic of a spinon excitation. The unpaired spin hops on the sites in a momentum eigenstate,
 while the valence bonds resonate among many configurations.}
\label{fig:spinon}
\end{center}
\end{figure}

Physically, the $z_\alpha$ operator creates a `spinon' excitation above the spin liquid state. This is a charge neutral, spin $S=1/2$ particle,
represented by a single unpaired spin in a background sea of resonating valence bonds: see Fig.~\ref{fig:spinon}.
In the formulation above, the spinon is a boson. Fermionic spinons are also possible, as we will discuss below in Section~\ref{sec:mott},
and appear in the present approach as bound states of spinons and visons.\cite{krs,sak,rc}

\subsection{Visons}
\label{sec:visons}

Visons are spinless, chargeless, excitations of a wide class of spin liquids. 
They can be viewed as the `dark matter' of condensed matter physics, being very
hard to detect experimentally despite (in many cases) carrying the majority of the entropy
and excitation energy.\cite{z2} They play a crucial role in delineating the structure of the excitations
and phase diagram of quantum antiferromagnets. 

\begin{figure}
\begin{center}
 \includegraphics[width=2.5in]{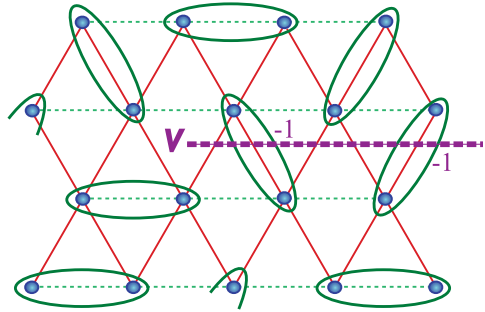}
 \caption{Schematic of a vison excitation. It is a vortex-like excitation in the spin liquid, created by inserting
 the indicated phase factors for each configuration of valence bonds}
\label{fig:vison}
\end{center}
\end{figure}

At the simplest level, a vison can be described \cite{rstl,rc} by the caricature of a wavefunction in Fig.~\ref{fig:vison}.
We choose an arbitrary `branch cut' extending from the center of a vison out to infinity, and insert a factor of (-1)
for each valence bond intersecting the branch cut. This yields a topological vortex-like excitation above the ground state.
The motion of the point $V$ in Fig.~\ref{fig:vison} in a momentum eigenstate yields a particle, which we represent by a 
real field $v$. 

Note that the particle $v$ is located on the lattice dual to the spins. An important aspect \cite{sf,jalabert} of its motion on 
the dual lattice is that it is moving in an average background flux: just as vortices in a superfluid experience the
background matter as an effective magnetic field (which is responsible for the magnus force), so do
visons experience a net flux of $\pi$ per direct lattice site containing a $S=1/2$ degree of freedom in the underlying
antiferromagnet. So we have to diagonalize the Hofstadter Hamiltonian of particles hopping on a lattice with flux to
obtain the proper vison eigenstates. This straightforward procedure has two important consequences:\cite{sf,jalabert,courtney,bbbss} \\
 ({\em i\/}) The vison
spectrum has a non-trivial degeneracy tied to the flux per unit cell. We denote the degenerate vison species by fields $v_a$,
where the index $a$ ranges over $1 \ldots N_v$, where $N_v \geq 2$ is the vison degeneracy.\\
({\em ii\/}) The vison eigenstates have non-trivial transformation properties under the space group of the lattice in Fig.~\ref{fig:tri}.
This transformation is connected to the structure of the wavefunction of the Hofstadter Hamiltonian and will
be important later in determining the nature of the phases proximate to the spin liquid.

\subsection{Solvable model}
\label{sec:kitaev}

Before writing down the field theory of the spinons, $z_\alpha$, and visons $v_a$ of the $S=1/2$ Heisenberg
spin model discussed above, we take a detour to describe an instructive exactly solvable model. This model
was introduced by Kitaev, \cite{kitaev} and is the simplest Hamiltonian containing spinon and vison-like degrees of freedom.
The solution of this model will make it clear that the spinons and visons are the electric and magnetic charges of
an underlying gauge theory.

\begin{figure}
\begin{center}
 \includegraphics[width=2in]{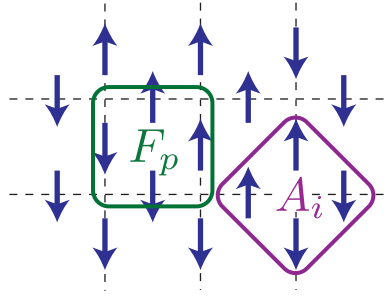}
 \caption{The two terms in the $H_K$ in Eq.~(\ref{ehk}).}
\label{fig:kit0}
\end{center}
\end{figure}

The Kitaev Hamiltonian can be written as
\begin{equation}
H_K = - J_1 \sum_i A_i - J_2 \sum_p F_p \label{ehk}
\end{equation}
where $i$ and $p$ extend over the sites and plaquettes of the square lattice, and $J_{1,2}$ are positive coupling
constants. The operators $A_i$ and $F_p$ are defined in terms $S=1/2$ Pauli spin operators ${\bm \sigma}_{\ell}$ which resides on the links, $\ell$,
of the square lattice; see Fig.~\ref{fig:kit0}. 
We have
\begin{equation}
A_i = \prod_{\ell \in \mathcal{N}(i)} \sigma^z_{\ell}
\end{equation}
where $\mathcal{N}(i)$ extends over the 4 links which terminate on the site $\ell$, and
\begin{equation}
F_p = \prod_{\ell \in p} \sigma^x_{\ell}
\end{equation}
where now the product is over the 4 links which constitute the plaquette $p$. The key to the solvability
of the Kitaev model is that these operators all commute with each other, as is easily checked
\begin{equation}
[A_i , A_j ] = [F_p, F_{p'}] = [A_i, F_p] = 0.
\end{equation}
Despite these seemingly trivial relations, the eigenstates have quite an interesting structure, as we will see.

\begin{figure}
\begin{center}
 \includegraphics[width=1.7in]{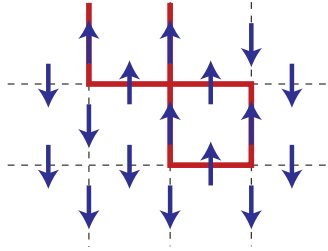}
 \caption{A component of the ground state of $H_K$. The red lines connect up spins and form closed loops}
\label{fig:kit1}
\end{center}
\end{figure}

\begin{figure}
\begin{center}
 \includegraphics[width=2.5in]{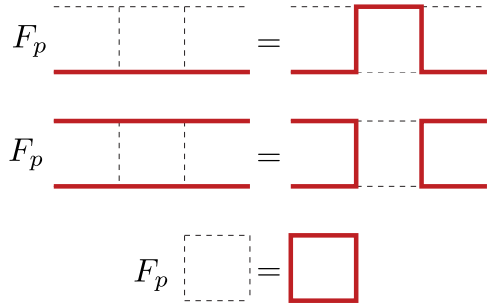}
 \caption{Action of the $F_p$ operator on sample configurations. Here $p$ is the center plaquette of the left hand configurations.}
\label{fig:kit2}
\end{center}
\end{figure}

The ground state is the unique state in which all the $A_i = 1$ and the $F_p = 1$. Let us write this state
in the basis of the $\sigma^z_\ell$ eigenstates. For $A_i = 1$ we need an even number of up (or down) spins
on the 4 links connected to each site $i$. If we now color each link with an up spin, this means that the terms
in the ground state consist only of closed loops of colored links: this is illustrated in Fig~\ref{fig:kit1}. 
Henceforth, we will
identify the states by their associated configuration of colored links. In this language, the action of $F_p$ on a plaquette
is to flip the color of each link in the plaquette; this has the consequence of moving, creating, and reconnecting the loops
in the ground state, as illustrated in Fig.~\ref{fig:kit2}. 
It is now evident that to obtain a state with eigenvalue $F_p=1$
for all $p$, we should simply take the equal positive superposition of all closed loop configurations on the square lattice.
This defines the spin liquid state in geometric terms.

Now let us describe the excited states. These turn out to be highly degenerate, an artifact of the solvable model.

\begin{figure}
\begin{center}
 \includegraphics[width=2.3in]{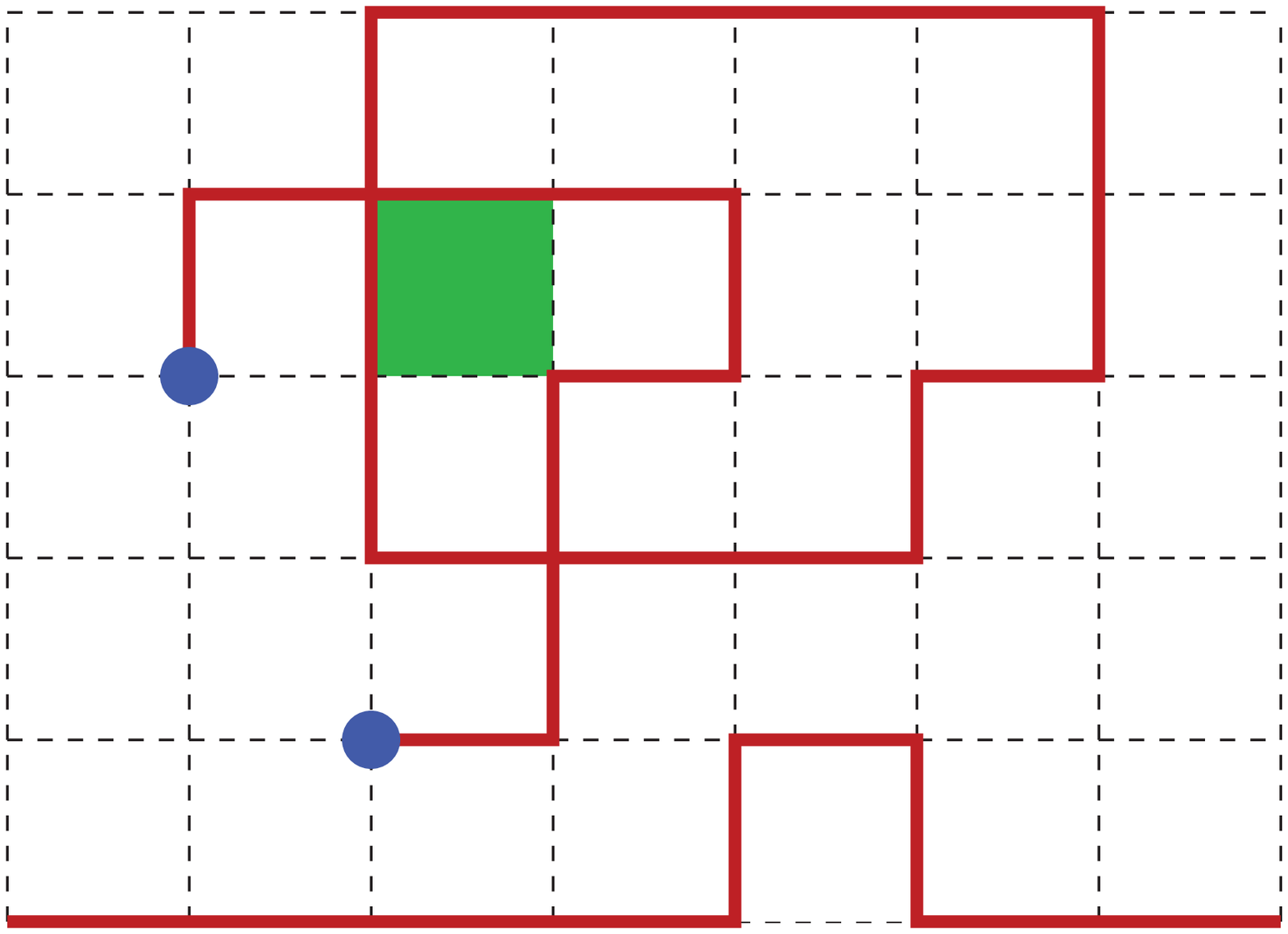}
 \caption{A component of a state with 2 spinons and 1 vison. The spinons are the blue circles at which the red lines
 end. The vison is the green plaquette on which $F_p$ has eigenvalue $-1$.}
\label{fig:kit3}
\end{center}
\end{figure}

The spinon excitation is a broken `bond' in the $\sigma^z_\ell$ basis. (Because there is no conserved spin quantum number here,
the spinon does not carry spin, but does disrupt the local exchange energy.) This is obtained by having a colored link end at a site $i$.
The spinon state has the eigenvalues $A_i = -1$ and $A_j = 1$ for all $j \neq 1$. We still retain $F_p=1$ on all plaquettes,
and so the spinon state is the equal superposition of all loop configurations on the lattice with a single free end at site $i$, as
shown in Fig.~\ref{fig:kit3}. 
If we interpret the Hamiltonian $H_K$ as a $Z_2$ gauge theory, then the spinon carries a $Z_2$ electric
charge.\cite{rstl,sr,wen1} Note also that the stationary spinon is an eigenstate, and so the spinon is infinitely massive. A generic model would
have spinons moving in momentum eigenstates, with a finite mass determining the spinon dispersion.

The vison has the complementary structure. It has $F_p=-1$ on a single plaquette $p$, and $F_{p'}=1$ for all $p' \neq p$.
It returns $A_i = 1$ for all sites $i$, and so can be described geometrically by closed loop configurations.
The wavefunction is still the superposition of all closed loop configurations, just as in the ground state. However, the signs
of some of the terms have been flipped; Starting with the loop-free configuration, each time a loop moves across the plaquette
$p$, we pick up a factor of $-1$ (see Fig.~\ref{fig:kit3}). 
In the $Z_2$ gauge theory language, the vison carries $Z_2$ magnetic flux.\cite{rstl,sf}
Again, the stationary vison is an eigenstate, but a more realistic model will have a vison with a finite mass.

By choosing $A_i = \pm 1$ and $F_p = \pm 1$ we can now easily extend the above constructions to states with
arbitrary numbers of spinons and visons. Indeed, these states span the entire Hilbert space of $H_K$. A typical state
is sketched in  Fig.~\ref{fig:kit3}. An examination of such states also reveals an important generic feature of the
dynamics of spinons and visons: when a spinon is transported around a vison (or vice versa) the overall wavefunction
picks up a phase factor of $(-1)$. In other words, spinons and visons are mutual {\em semions\/}.

Kitaev's construction generalizes to a large number of solvable models, some with much greater degrees of 
complexity. \cite{rk,sondhi,pasquier,wen2,freedman,stringnet,kitaev2,vidal,yao}
The quasiparticles of these models carry electric and magnetic charges of a variety of gauge groups, and in
some cases obey non-Abelian statistics.

\subsection{Field theory of spinons and visons}
\label{sec:cs}

Let us now return to the class of $S=1/2$ Heisenberg antiferromagnets considered in Section~\ref{sec:tri}.
There are two key differences from the solvable Kitaev model: ({\em i\/}) the spinons, $z_\alpha$, carry 
a global SU(2) spin label $\alpha$, and ({\em ii\/}) the visons $v_a$ have an additional flavor label, $a$,
associated with their non-trivial transformation under the lattice space group. The visons of the Kitaev model
do not have a flavor degeneracy because they do not move in an average background flux, a consequence
of there being an even number of $S=1/2$ spins per unit cell in this solvable model. However, the mutual
semionic statistics of the spinons and visons does extend to the Heisenberg antiferromagnets, and can be implemented
in a Chern-Simons field theory for its excitations.

In many of the interesting cases, including the lattice in Fig.~\ref{fig:tri}, it is possible to combine the real vison
fields $v_a$ into complex pairs, and coupling the resulting fields consistently to a U(1) gauge field. \cite{freedman,sondhibf,wenlevin,cenke}
For the lattice in Fig.~\ref{fig:tri}, the simplest possibility is \cite{cenke} that there are only 2 vison fields,
and these combine into a single complex vison $V=v_1 + i v_2$. Coupling this vison to a U(1) gauge field, $b_\mu$,
we then have the proposed field theory \cite{cenke} for Heisenberg antiferromagnets on the lattice in Fig.~\ref{fig:tri}
\begin{eqnarray}
\mathcal{S}_{zv} &=& \int d^2 r d\tau \Big\{ \mathcal{L}_z + \mathcal{L}_v + \mathcal{L}_{cs} \Bigr\} \nonumber \\
\mathcal{L}_z &=& |(\partial_\mu - i
a_\mu)z_\alpha|^2  + s_z |z_\alpha|^2  + u_z (|z_\alpha|^2)^2 \nonumber \\
\mathcal{L}_v &=&  |(\partial_\mu - i b_\mu)V|^2 + s_v |V|^2  +
u_v |V|^4 \nonumber \\
\mathcal{L}_{cs} &=&
\frac{ik}{2\pi} \epsilon_{\mu\nu\lambda}a_\mu\partial_\nu
b_\lambda , \label{lcs}
\end{eqnarray}
where $\mu$ is a spacetime index, and we need the Chern-Simons term $\mathcal{L}_{cs}$ at level $k=2$. 
This field theory replaces the Landau-Ginzburg field theory $\mathcal{S}_{LG}$ in Eq.~(\ref{slg})
for quantum `disordering' magnetic order in antiferromagnets with one $S=1/2$ spin per unit cell.
Now we have two tuning parameters, $s_z$ and $s_v$, and these yield a more complex phase diagram, \cite{cenke}
to be discussed shortly. We note in passing that the theory in Eq.~(\ref{lcs}) bears a striking resemblance to supersymmetric
gauge theories \cite{m2d} much studied in recent years because of their duality to M theory on $AdS_4 \times
S^7/Z_k$: in both cases we have doubled Chern-Simons theories with bifundamental matter.

Apart from the usual N\'eel order parameter ${\bm \varphi}$ defined in Eq.~(\ref{neel}), the presence of the vison field
$V$ allows us to characterize other types of broken symmetry. The space group transformation properties of $V$ show
that \cite{courtney,bbbss,cenke} $V^2$ is the VBS order parameter characterizing the broken lattice symmetry of
the state in Fig.~\ref{fig:vbs}. Secondly, in the phases where the vison $V$ is gapped ($s_v > 0$), we can freely integrate the $b_\mu$ gauge field,
and the Chern Simons term in Eq.~(\ref{lcs}) has the consequence of quenching $a_\mu$ to a $Z_k$ gauge field.
In such phases, the composite field $z_\alpha z_\beta$ is gauge invariant and can also be used to characterize broken symmetries;
for the case being considered here, this order parameter characterizes an antiferromagnetic 
state with spiral spin order.

\begin{figure}
\begin{center}
 \includegraphics[width=4.5in]{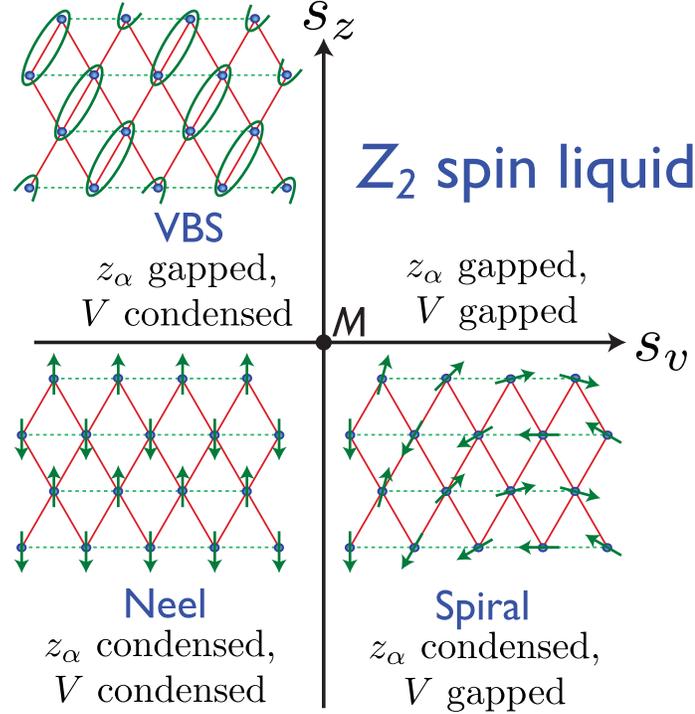}
 \caption{A phase diagram for the Heisenberg antiferromagnet on the lattice in Fig.~\ref{fig:tri} obtained \cite{cenke} from
 the field theory $\mathcal{S}_{zv}$ in Eq.~(\ref{lcs}). The same phase diagram was obtained \cite{sr} earlier by other methods.}
\label{fig:phase}
\end{center}
\end{figure}

Using these considerations, the field theory in Eq.~(\ref{lcs}) leads to the schematic phase diagram \cite{rstl,sr,cenke}
shown in Fig.~\ref{fig:phase}.
The phases are distinguished by whether one or more of the spinon and vison fields condense.
Apart from the phases with broken symmetry which we have already mentioned (the N\'eel, spiral, and VBS states),
it contains a spin liquid state with no broken symmetry. This is called a $Z_2$ spin liquid \cite{rstl,wen1} because the spinons
and visons carry only a $Z_2$ quantum number, a consequence of the arguments in the previous paragraph.

Note that this theoretical phase diagram contains the N\'eel and VBS states found in the experimental phase
diagram in Fig.~\ref{fig:kato}. It is possible that some of the compounds with magnetic order with larger values of $J'$
actually have spiral order---neutron scattering experiments are needed to fully characterize the magnetic order.

The series expansion study of Weihong {\em et al.} \cite{wms} of the nearest neighbor antiferromagnet on the lattice
in Fig.~\ref{fig:tri} finds
the N\'eel, VBS, and spiral phases, and indicates that the point $J' = J$ is not too far from the multicritical point M
in Fig.~\ref{fig:phase}.  This suggests that we analyze spin liquid compounds like X=EtMe$_3$Sb in Fig.~\ref{fig:kato}
by using a field theory of quantum fluctuations close to M; this point of view is discussed further below in Section~\ref{sec:triangle}.

Another notable feature of Fig.~\ref{fig:phase} is the presence of a direct transition between phases which
break distinct symmetries---this is the transition between the N\'eel and VBS states. Such a direct transition was
discussed in some detail in early work, \cite{rsl,rsb} where it was shown that the VBS order appeared as a consequence
of Berry phases carried by hedgehog/monopole tunneling events which proliferated in the non-N\'eel phase.
Here, we have given a different formulation of the same transition in which the Berry phases were associated instead
with visons. VBS order has now been observed
proximate to the N\'eel phase in numerical studies on a number of Heisenberg antiferromagnets
on the square, \cite{sandvik,cirac,kawashima,lauchli} triangular, \cite{wms} checkerboard, \cite{chung,fouet,tchern,sfb,poilblanc} and honeycomb lattices. \cite{fouet2} We have also noted earlier the experimental detection of VBS-like correlations \cite{kohsaka1,kohsaka2}
in underdoped cuprates which are also proximate to the N\'eel state.
 
A direct second-order N\'eel-VBS
transition is forbidden in the Landau-Ginzburg framework, except across a multicritical point.  
Arguing that transitions that violate this framework are possible at quantum critical points, 
Senthil {\em et al.}\cite{senthil} proposed a field theory for the N\'eel-VBS transition based upon
the Berry phase-induced suppression of monopoles at the critical point;  so the criticality  was
expressed in a monopole-free theory. \cite{mv}
In the approach reviewed above, this critical theory is
obtained from $\mathcal{S}_{zv}$ in Eq.~(\ref{lcs}) by condensing $V$ and ignoring the gauge field
$b_\mu$ which is now `Higgsed' by the $V$ condensate; the resulting theory is $\mathcal{L}_{z}$, 
the CP$^1$ model of $z_\alpha$ and the U(1) gauge field $a_\mu$.
A number of large-scale computer studies \cite{sandvik,melkokaul,wiese,kuklov,kuklov2,mv2,flavio} have examined the N\'eel-VBS transition.
The results provide strong support for the suppression of monopoles near the transition, and 
for the conclusion that the CP$^1$ field theory properly captures
the low energy excitations near the phase transition.\cite{sandvik,flavio,ssnat} However, some simulations \cite{wiese,kuklov,kuklov2}
present evidence for a weakly first-order transition, and this could presumably be a feature
of the strong-coupling regime of the CP$^1$ field theory.

\section{Spin liquids near the Mott transition}
\label{sec:mott}

We now turn to the second route to exotic insulating states outlined in Section~\ref{sec:intro}. Rather than using the insulating spin
model in Eq.~(\ref{HJ}), we include the full Hilbert space of the Hubbard model in Eq.~(\ref{HU}), and begin with a conventional
metallic state with a Fermi surface. The idea is to turn up the value of $U$ at an odd-integer filling of the elctrons
so that there is a continuous transition to an insulator
in which a `ghost' Fermi surface of neutral fermionic excitations survives.\cite{florens}

This idea is implemented by writing the electron annihilation operator as a product of a charged boson, $b$, and a neutral spinful fermion 
$f_\alpha$ (the spinon)
\begin{equation}
c_\alpha = b f_\alpha. \label{cbf}
\end{equation}
Then we assert that with increasing $U$ the boson undergoes a superfluid-to-Mott insulator transition just as
in a Bose Hubbard model; this is possible because the bosons are spinless and at an odd-integer filling. 
The superfluid phase of the bosons is actually a metallic Fermi liquid state for the physical electrons:
we see this from Eq.~(\ref{cbf}), where by replacing $b$ by its c-number expectation value $\langle b \rangle$, the $f_\alpha$
acquire the same quantum numbers as the $c_\alpha$ electrons, and so the $f_\alpha$ Fermi surface describes a conventional metal.
However, the Mott insulator for the bosons is also a Mott insulator for the electrons, with a gap to all charged excitations.
Under suitable conditions, the $f_\alpha$ Fermi surface survives in this insulator, and describe a continuum a gapless, neutral
spin excitations---this is the spinon Fermi surface.

The most  complete study of such a transition has been carried out on the honeycomb lattice at half-filling.\cite{leelee,hermele} In this
case the metallic state is actually a semi-metal because it only contains gapless electronic excitations at isolated Fermi points
in the Brillouin zone (as in graphene). The electronic states near these Fermi points have a Dirac-like spectrum, and the
use of a relativistic Dirac formalism facilitates the analysis. The low energy theory for the neutral 
Dirac spinons, $\Psi$ in the insulating phase has the schematic form
\begin{equation}
\mathcal{S}_D = \int d^2 r d\tau \Bigl\{ \overline{\Psi}  \gamma^\mu (\partial_\mu - i a_\mu ) \Psi \Bigr\} \label{asl}
\end{equation}
where $\gamma^\mu$ are the Dirac matrices in 2+1 dimensions, and $a_\mu$ is an emergent gauge field associated with
gauge redundancy introduced by the decomposition in Eq.~(\ref{cbf}). Depending upon the details of the lattice implementation,\cite{hermele}
$a_\mu$ can be a U(1) or a SU(2) gauge field. For a large number of flavors, $N_f$, of the Dirac field (the value of $N_f$ is determined
by the number of Dirac points in the Brillouin zone), the action $\mathcal{S}_D$ is known to describe a conformal field theory (CFT).
This is a scale-invariant, strongly interacting quantum state, with a power-law spectrum in all excitations, and no well-defined
quasiparticles. In the present context, it has been labeled an algebraic spin liquid.\cite{rantner}

Closely related algebraic spin liquids have also been discussed on the square \cite{am,ioffe,rantner,mother,alicea} 
and kagome \cite{hastings,ran,ran2} lattices. In these cases, the bare lattice dispersion of the fermions does
not lead to a Dirac spectrum. However, by allowing for non-zero average $a_\mu$ fluxes on the plaquettes, and optimizing these
fluxes variationally, it is found that the resulting `flux' states do often acquire a Dirac excitation spectrum.

One of the keys to the non-perturbative 
stability of these algebraic spin liquids \cite{ioffe,mother,alicea} is the suppression of tunneling events associated
with monopoles in the $a_\mu$ gauge field. This has so far only been established in the limit of large $N_f$.
For the N\'eel-VBS transition discussed at the end of Section~\ref{sec:cs}, there is now quite good evidence
for the suppression of monopoles near the transition.\cite{sandvik,ssnat}. For the present fermionic algebraic
spin liquids, the main numerical study is by Assaad \cite{assaad} on the square lattice for antiferromagnets with global SU($N$)
symmetry; he finds evidence for an algebraic spin liquid for SU(4) but not for SU(2). The results of Alicea \cite{alicea} indicate
that single monopole tunnelling events are permitted on the square lattice, and these are quite likely to be relevant perturbations
away from the fermionic algebraic spin liquid.

On the triangular lattice, an analysis along the above lines has been argued to lead to a genuine Fermi surface
of spinons.\cite{mot} In this case, the suppression of the monopoles is more robust,\cite{mother,herm2,sslee}
but `$2k_F$' instabilities of the Fermi surface could lead to ordering at low temperatures. \cite{aim,palee}
A detailed study of the finite temperature crossovers near a postulated continuous metal-insulator transition has been 
provided for this case. \cite{senthilmit}

An interesting recent numerical study \cite{sheng} has presented evidence of the remnant of a spinon Fermi surface
in spin ladders: the spectrum of a `triangular' ladder contains excitations that can be identified with spinon Fermi points,
and these can be regarded as remnants of a spinon Fermi surface after quantizing momenta by periodic boundary
conditions in the transverse direction. It will be interesting to see if the number of such Fermi points increase as more
legs are added to the ladder, as is expected in the evolution to a Fermi surface in two dimensions.

\section{Exotic metallic states}

This section will consider extend the ideas of Section~\ref{sec:tri} from insulating to conducting states.
This will yield metals with Fermi surfaces, some or all of whose quasiparticles do not have traditional charge $\pm e$ and
spin $S=1/2$ quantum numbers, and the temperature dependencies of various thermodynamic
and transport co-efficients will differ from those in traditional Fermi liquid theory.
In keeping with the unifying strategy outlined in Section~\ref{sec:intro}, we will begin with a conventional Fermi liquid
state, and induce strong quantum fluctuations in its characteristic `quantum order'. We will consider the breakdown
of Kondo screening in Section~\ref{sec:ffl}, and a quantum fluctuating metallic spin density wave in Section~\ref{sec:acl}.

\subsection{Fractionalized Fermi liquids}
\label{sec:ffl}

We begin with the heavy Fermi liquid state, usually obtained from the Kondo-Heisenberg model.
The latter is derived from a two-orbital Hubbard model, $H_U$, in which the repulsive energy $U$ associated with 
one of the orbitals (which usually models $f$ orbitals in intermetallic compounds) is much larger than that is the second
orbital (representing the conduction electrons). 
In such a situation, we can perform a canonical transformation to a reduced Hilbert space in which
the charge on the $f$ orbital is restricted to unity, and its residual spin degrees of freedom
are represented by a $S=1/2$ spin operator ${\bf S}_i$. These couple to each other and the conduction electrons
in the Kondo-Heisenberg Hamiltonian
\begin{equation}
H_{KH}  = \sum_{i<j} J_{ij} {\bf S}_i \cdot {\bf S}_j 
+ \sum_{k} \varepsilon(k) c_{k \alpha}^\dagger c_{k \alpha} + \frac{J_K}{2} \sum_{i} {\bf S}_i \cdot c_{i\alpha} {\bf \sigma}_{\alpha\beta} c_{i \beta}
\end{equation}
where conduction electrons with dispersion $\varepsilon(k)$ 
are annihilated by $c_{k \alpha}$ and $c_{i \alpha}$ in momentum and real space respectively.

The heavy Fermi liquid state is obtained for $J_K \gg |J_{ij}|$ where the ${\bf S}_i$ are predominantly Kondo screened
by formation of singlets with the conduction electrons. The structure of this state is most easily revealed by writing
the local moments in terms of neutral fermionic spinons ${\bf S}_i = f_{i \alpha}^{\dagger} {\bm \sigma}_{\alpha\beta} f_{i \beta}$.
Then Kondo screening can be identified with the condensation of the bosonic field $B \sim f_{\alpha}^\dagger c_{\alpha}$.
The $f$ and $c$ fermions strongly hybridize in the resulting state, leading to a ``large'' Fermi surface of the composite
fermionic quasiparticle: the volume enclosed by this Fermi surface counts both the $f$ and $c$ fermions, and so obeys
the traditional Luttinger rule. The absence of a bare hopping matrix element for the $f$ fermions is responsible for the heavy mass
of the quasiparticles at the Fermi surface.

Now consider increasing the values of the Heisenberg exchange interactions $J_{ij}$. We do this \cite{burdin} 
with a set of $J_{ij}$ which on
their own would prefer to form one of the spin liquid states discussed in Section~\ref{sec:tri}. 
Eventually, it will be preferable for the $f$ moments to form singlets with each other, rather than being
screened by the conduction electrons. This breakdown of Kondo screening happens at a sharp phase transition \cite{ssv,svs} in 
an effective gauge theory describing the disappearance of the Higgs condensate of $B$. Across the transition,
we obtain a new non-Fermi liquid state labeled the `fractionalized Fermi liquid'. The structure of this state
is easily understood by adiabatic continuation from the $J_K=0$ limit of $H_{KH}$: the $f$ moments form one
of the spin liquid states of Section~\ref{sec:tri}, while the $c$ conduction electrons form a Fermi surface of 
Landau-like, charge $\pm e$, spin $S=1/2$ quasiparticles on their own. The unusual property of this decoupled limit
violates the traditional Luttinger rule on the Fermi surface volume, which now includes only `small' number of conduction
electrons. The existence of this small Fermi surface is intimately linked to the presence of an exotic spin liquid 
on the $f$ sites. It is the stability of this spin liquid which ensures adiabatic continuity, and retains the small Fermi volume
even when $J_K$ is non-zero.\cite{ssv,svs} The observable properties of this phase, of its transition to the
heavy Fermi liquid state, and the connections to experiments on correlated electron compounds
have been discussed elsewhere.\cite{svs,coleman,paul}

\subsection{Algebraic charge liquids}
\label{sec:acl}

We turn here to the last of the methods noted in Section~\ref{sec:intro} for obtaining exotic states.
We will begin with antiferromagnetically ordered state, but in a Fermi liquid, rather than in an insulator.
It is traditional to refer to such ordered metals as spin density wave (SDW) states. Just as in the insulator,
the SDW order can be characterized by a vector order parameter ${\bm \varphi}$. We can describe the loss of this
SDW order in the conventional Landau-Ginzburg framework: this leads to a non-magnetic Fermi liquid
state, and we will review this theory below. However, we wish to consider a possibility here which is the analog
of that described in Section~\ref{sec:tri} for insulating antiferromagnets: we characterize the SDW order not
by the vector ${\bm \varphi}$ but by the bosonic spinor $z_\alpha$ in Eq.~(\ref{neel}). We wish to describe the loss of SDW
order by a gauge theory of $z_\alpha$ fluctuations. This will lead to a non-Fermi liquid state \cite{rkk1,rkk2,rkk3,gal} labeled an algebraic
charge liquid (ACL). Just as was the case for insulators in Section~\ref{sec:cs}, consistency of a $z_\alpha$ plus gauge
field description of such a quantum critical point requires suppression of hedgehogs in the ${\bm \varphi}$ field
(which are monopoles in the $a_\mu$ field).
This is facilitated here by the presence of Fermi surfaces which strongly suppress monopoles \cite{herm2,sslee},
and we don't have to appeal to delicate Berry phase cancellations which were needed in the insulator.

\begin{figure}
\begin{center}
 \includegraphics[width=5in]{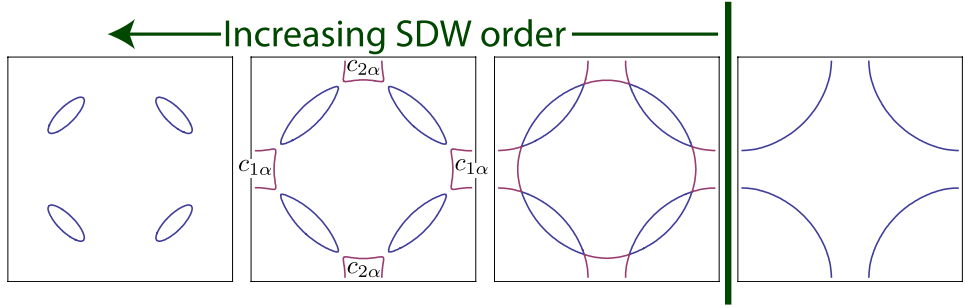}
 \caption{Evolution of the Fermi surface of the hole doped cuprates in a conventional SDW theory
 as a function of the magnitude of the SDW order $|{\bm \varphi}|$. The right panel
is the large Fermi surface state with no SDW order. The onset of SDW order induces the formation of electron  
(red) and hole (blue) pockets. With further increase of $|{\bm \varphi}|$, the electron pockets disappear and only 
hole pockets remain.
 }
\label{fig:sdw}
\end{center}
\end{figure}

It is useful to begin with a review of the Landau-Ginzburg approach to the loss of SDW order in a metal.
The mean-field theory evolution of the Fermi surface with increasing SDW order is shown in Fig.~\ref{fig:sdw},
for a Fermi surface configuration appropriate for the hole-doped cuprates.\cite{morr} 
Right at the quantum critical point,
the SDW fluctuations, ${\bm \varphi}$ 
connect points on the SDW Fermi surface, and so can decay into a large density of states
of particle-hole excitations. The damping induced by this particle-hole continuum modifies the effective action
for ${\bm \varphi}$ from Eq.~(\ref{slg}) by adding a strongly relevant term: \cite{maki,hertz}
\begin{equation}
\mathcal{S}_H = \mathcal{S}_{LG} + \int \frac{d^2 k}{4 \pi^2} \int \frac{d \omega}{2 \pi} |\omega| |{\bm \varphi} (k, \omega)|^2
\end{equation}
where $\omega$ is a Matsubara imaginary frequency. The theoretical and experimental implications of this modified 
Landau-Ginzburg theory
have been explored extensively in the literature. From Fig.~\ref{fig:sdw} we see that this theory 
describes a quantum transition from a SDW state with electron and hole pockets, to a Fermi liquid state with a large Fermi surface.

Now we turn to the $z_\alpha$ description of the loss of SDW order. For this, we begin from the second panel in Fig.~\ref{fig:sdw},
the Fermi liquid SDW ordered state with electron and hole pockets. We introduce a parameterization of the electronic
excitations in which their spin polarization is quantized along the direction of the local SDW order, determined by the local
orientation of the $z_\alpha$. Thus, let $c_{1\alpha}$ represent the fermionic quasiparticles in one of the electron pockets (see
Fig.~\ref{fig:sdw}).
We express these electrons in terms of fermions $g_\pm$ whose spin is polarized along the direction of the local
SDW order:
\begin{equation}
\left( \begin{array}{c} c_{1\uparrow} \\ c_{1\downarrow} \end{array} \right) = \mathcal{R}_z \left( \begin{array}{c} g_+ \\ g_- \end{array} \right)~~~~~;~~~~~\mathcal{R}_z \equiv \left( \begin{array}{cc}
z_\uparrow & -z_\downarrow^\ast  \\
z_\downarrow &  z_\uparrow^\ast   
 \end{array} \right)
 \label{zg1}
\end{equation}
For a uniformly polarized SDW state along the $\hat{z}$ direction, we have $(c_{1\uparrow} , c_{1\downarrow}) = (g_+ ,  g_-)$, and so
the $g_\pm$ are the usual up and down spin electron operators.
However, Eq.~(\ref{zg1}) allows us to describe an arbitrary spacetime dependent orientation of the SDW order by using a `rotating reference frame' defined by the SU(2) rotation matrix $\mathcal{R}_z$.
A similar parameterization applies to fermions in the electron pocket which is shifted from the above pocket by the SDW ordering
wavevector $(\pi,\pi)$; we denote these fermions $c_{2\alpha}$ (see
Fig.~\ref{fig:sdw}). For a uniform SDW order along the $\hat{z}$ direction, the Hartree-Fock theory of the mixing between the electron
eigenstates now shows that $(c_{2\uparrow} , c_{2\downarrow}) = (g_+ , - g_-)$; generalizing this to an arbitrary orientation
as in Eq.~(\ref{zg1}), we have \cite{rkk3,gal}:
\begin{equation}
\left( \begin{array}{c} c_{2\uparrow} \\ c_{2\downarrow} \end{array} \right) = \mathcal{R}_z \left( \begin{array}{c} g_+ \\ -g_- \end{array} \right). \label{zg2}
\end{equation}
A similar analysis can also be carried out near the hole pockets.

We can now write down the general structure of the effective action controlling the $z_\alpha$ and $g_\pm$ modes.
Within the SDW ordered phase with $\langle {\bm \varphi} \rangle \neq 0$, 
this theory will be entirely equivalent to the conventional SDW Hartree-Fock theory,
which is just re-expressed here in a different set of fields. However, it will lead to a new exotic ACL phase in the 
state without SDW order, $\langle {\bm \varphi} \rangle = 0$. Constraints from gauge invariance and lattice symmetries
lead to the action \cite{leeholon,rkk1,rkk2,rkk3,gal}
\begin{eqnarray}
\mathcal{S}_{ACL} = \int d^2 r d\tau \Biggl\{\mathcal{L}_z &+&
 g_+^\dagger \left( \frac{\partial}{\partial \tau} + i a_\tau - \frac{1}{2m} ( \vec{\nabla} + i \vec{a} )^2 
- \mu \right) g_+ \nonumber \\
&+& g_-^\dagger \left( \frac{\partial}{\partial \tau} - i a_\tau - \frac{1}{2m} ( \vec{\nabla} - i \vec{a} )^2 
-\mu \right) g_- \Biggr\}. \label{acl}
\end{eqnarray}
The spinon component, $\mathcal{L}_z$, of $\mathcal{S}_{ACL}$ is the same as that in Eq.~(\ref{lcs}), and the parameter $s_z$ tunes
the theory from the SDW phase with $z_\alpha$ condensed, to the ACL with $z_\alpha$ gapped. The $g_\pm$ fermions
carry opposite charges under the $a_\mu$ gauge field, as is clear from the requirement that the $c_{1\alpha}$ and $c_{2\alpha}$
electron operators in Eqs.~(\ref{zg1}) and (\ref{zg2}) be gauge invariant.

An analysis of the SDW-ACL critical point has been carried out using Eq.~(\ref{acl}), and the result is relatively simple \cite{rkk3,senthilmit}: the $g_\pm$ fermions serve to damp the $a_\mu$ gauge field in a manner that it no longer
couples efficiently to the $z_\alpha$. At the critical point, the spin excitations are described by a theory of the
$z_\alpha$ alone, and we can ignore their coupling to the gauge field and the fermions: consequently, the critical
point is in the O(4) universality class, arising from the four real components of the $z_\alpha$.

The structure of the ACL phase is also clear from Eq.~(\ref{acl}). While the spinons are gapped, there are gapless charged excitations associated with Fermi surfaces of the spinless,\cite{rkk1,rkk2,rkk3,gal,wenholon,leeholon,shankar,ioffew,paul2}
 charge $-e$ carriers $g_\pm$.
A detailed application of this structure to the peculiar properties of the underdoped cuprates
has been discussed recently, \cite{gal} motivated by the evidence for the existence of electron pockets at high magnetic fields.\cite{louis} The attractive gauge force between the $g_\pm$ pockets causes them
to strongly pair into a $s$-wave paired state, with the local pairing $\langle g_+ g_- \rangle \neq 0 $ and independent of 
momentum on the Fermi surface. Remarkably, application of the relations in Eqs.~(\ref{zg1}) and (\ref{zg2})
shows that such a state actually corresponds to $d$-wave pairing of the physical electrons: \cite{rkk3} 
factorizing the expectation values of the fermions and bosons, we have
\begin{equation}
\left\langle c_{1 \uparrow} c_{1 \downarrow} \right\rangle = - \langle c_{2 \uparrow} c_{2 \downarrow} \rangle = 
\left\langle ( |z_\uparrow |^2 + |z_\downarrow |^2 ) \right\rangle \, \left\langle g_+ g_- \right\rangle.
\end{equation}
The paired electron
pockets also induce a weak proximity-pairing of the hole pockets, in a manner which is consistent with the overal $d$-wave
pairing symmetry of the electrons; this leads to gapless nodal fermionic excitations along the Brillouin zone
diagonals. These features have been proposed as an explanation of the `nodal-anti-nodal dichotomy'
observed in the underdoped cuprates.\cite{gal}

\section{Experiments on Mott insulators}
\label{sec:exp}

This concluding section will highlight recent experiments on a variety of Mott insulators. We will complement the discussion
by initially describing numerical studies of quantum spin models on the corresponding lattices. 
We will restrict our attention here to $S=1/2$ antiferromagnets; there are also interesting examples of 
exotic states in the higher spin insulators  FeSc$_2$S$_4$ \cite{chenbalents,buttgen} and NiGa$_2$S$_4$.\cite{naka}

\subsection{Triangular lattice}
\label{sec:triangle}

The most extensive numerical studies on the triangular lattice antiferromagnet have been carried out by the
Paris group.\cite{misguich1,misguich2,lhuillier} They have examined the phase diagram as a function of the
ratio $J_4/J_2$, where $J_2$ is the conventional 2 spin exchange as in Eq.~(\ref{HJ}), and $J_4$ is the 4-spin
ring exchange around all rhombi of the triangular lattice. For small $J_4/J_2$ there is convincing evidence \cite{chernyshev} 
for 3-sublattice antiferromagnetic order (this is a commensurate version of the `spiral' state in Fig.~\ref{fig:phase}).
For large $J_4/J_2$, there is reasonable evidence for a gapped $Z_2$ spin liquid state, similar to that
discussed in Section~\ref{sec:cs}, and in Fig.~\ref{fig:phase}. The situation for intermediate $J_4/J_2$ seems unresolved
at present system sizes. The theory of Fig.~\ref{fig:phase} would predict a direct second-order transition between these
phases, but it has also been argued that there is an intermediate spinon Fermi surface state. \cite{mot}

Turning to experiments, the most detailed early experiments were carried out by Coldea and collaborators \cite{radu1,radu2}
on Cs$_2$CuCl$_4$. This compound has the geometry of Fig.~\ref{fig:tri}, but with $J'/J$ large. Notice that in the limit 
$J'/J \rightarrow \infty$, this lattice becomes equivalent to a set of decoupled one-dimensional spin chains.
The neutron scattering experiments show spiral magnetic order, as is expected from the classical
ground state for large $J'/J$. However, they also show an anomalous and strong continuum of non-spin-wave excitations at
higher energies. Recent studies \cite{sb1,sb2} have argued that these anomalous excitations can be 
quantitatively explained in a theory which begins with the decoupled spin chain solution and includes the effects
of the inter-chain coupling, $J$, perturbatively.

We have already mentioned in Section~\ref{sec:tri} the extensive studies on the organic Mott insulators X[Pd(dmit)$_2$]$_2$
and their phase diagram in Fig.~\ref{fig:kato}. A closely related set of experiments have been 
carried out by Kanoda and collaborators \cite{kanoda0,kanoda1,kanoda2} on the organic compounds
$\kappa$-(ET)$_2$X. So far, the Mott insulator has been studied only
for X = Cu$_2$(CN)$_3$ and X=Cu[N(CN)$_2$]Cl (unlike the many more examples
in Fig.~\ref{fig:kato} for X[Pd(dmit)$_2$]$_2$). These compounds also have $S=1/2$ moments
on the lattice of Fig.~\ref{fig:tri}. The compound X=Cu[N(CN)$_2$]C
has $J'/J \approx 0.5$, and, as can be expected by analogy from Fig.~\ref{fig:kato},
it is clearly observed to have antiferromagnetic order.
Much attention has focused recently on the compound with X = Cu$_2$(CN)$_3$
which is quite close to the isotropic limit $J'/J \approx 1$, and is a candidate spin liquid
with no observed ordering at low $T$. Its properties are similar to 
the case X=EtMe$_3$Sb in Fig.~\ref{fig:kato}, and the analogy between these two series of compounds
would suggest a description of $\kappa$-(ET)$_2$  Cu$_2$(CN)$_3$ using the proximity to the quantum 
phase transitions in Fig.~\ref{fig:phase}; such a proposal has recently been examined in some detail. \cite{z2}
Taken in isolation with studies of $\kappa$-(ET)$_2$X, such a proposal appears to require
fine-tuning to place $\kappa$-(ET)$_2$  Cu$_2$(CN)$_3$ near a quantum phase transition,
but the analogy with the phase diagram of the X[Pd(dmit)$_2$]$_2$ in Fig.~\ref{fig:kato} makes
this proposal more natural. Also, we have noted earlier evidence from series expansion studies \cite{wms}
that the isotropic point $J'/J = 1$ is not too far from the point M in Fig.~\ref{fig:phase}.

Conflicting evidence on the nature of the low temperature state of $\kappa$-(ET)$_2$  Cu$_2$(CN)$_3$
has appeared in two recent experiments.\cite{kanoda2,yamashita}
Specific heat measurements \cite{kanoda2} show a non-zero value in the low temperature
extrapolation of $\gamma = C_P/T$, which would be consistent with a spinon
Fermi surface.\cite{palee} However, the extraction of the electronic
specific heat requires the subtraction of a large nuclear contribution. This subtraction
has been questioned by Yamashita {\em et al.} \cite{yamashita}, who instead measured
the thermal conductivity, $\kappa$. This is not subject to contamination by a nuclear
contribution, and they found a zero extrapolation of $\kappa/T$ as $T \rightarrow 0$.
Indeed, their low $T$ behavior was consistent the thermal transport via
a gapped electronic excitation. Qi {\em et al.}\cite{z2,cenke} have proposed identification
of this gapped excitation with the vison, and have argued that this yields a consistent
explanation of a variety of experiments.

\subsection{Kagome lattice}
\label{sec:kagome}

The nearest neighbor antiferromagnet on the kagome lattice has been examined by a
variety of numerical studies. The most recent evidence \cite{kag3,kag4,kag5} points consistently
to a ground state with a spin gap of $0.05J$ and VBS order. The pattern of the VBS order
is quite complex, with a large unit cell, but was anticipated in studies based upon the $1/N$ expansion
of the SU($N$) antiferromagnet. \cite{rs1,kag1,kag2}

Turning to experiments, recent examples of $S=1/2$ kagome antiferromagnets 
are found in the compounds \cite{kindo1,kindo2} A$_2$Cu$_3$SnF$_{12}$ with A=Cs and Rb.
The A=Cs compound has the perfect kagome structure at room temperature,
but undergoes a structural transition at $T=185$ K and  details of the low temperature structure are
not yet clear; the system orders magnetically $T=20$ K. The A=Rb compound is a distorted kagome
already at room temperature, which leads to distinct exchange interactions between nearest neighbor pairs with
an average of
$J \approx 200$K. This system does not order magnetically but has a spin gap
of 21 K. It is tempting to associate the structural distortion and the spin gap with formation of the
VBS state---a recent analysis along these lines has been provided by Yang and Kim.\cite{yangkim}

Volborthite Cu$_3$V$_2$O$_7$(OH)$_2 \cdot$2H$_2$O is another interesting kagome compound.\cite{vol1,vol2,vol3} Here all the nearest neighbor exchange constants appear equal, and magnetic order of the spins
is observed, albeit with a significant amount of spatial randomness.

Yamabe {\em et al.} \cite{tanaka} studied the compounds Cs$_2$Cu$_3$MF$_{12}$, with M=Zr and Hf, which form single crystal
$S=1/2$ kagome antiferromagnets with large exchange constants, $J = 360$ and $540$ K respectively.
These undergo structural transitions (to a not yet determined structure) at $T= 210$ and $175$ K respectively. The low transition temperatures (compared to $J$) suggest that exchange interactions play a role here, and that there is 
connection between the structural transition and the physics of VBS ordering.
At lower temperatures, magnetic ordering is observed at $T = 23.5$ and $24.5$ K.
Dzyaloshinsky-Moriya (DM) interactions are allowed on the kagome lattice \cite{dmkag},
and Yamabe {\em et al.} noted that these are likely the driving force for the magnetic ordering.

Finally, much attention has focused on the $S=1/2$ compound herbertsmithite
ZnCu$_3$(OH)$_6$Cl$_2$. This has $J \approx 170$ K and 
no observed ordering or structural distortion. \cite{helton,ofer,mendels,imai}
However, there is an appreciable amount of substitutional disorder between the Zn and Cu sites
which affects the low $T$ behavior. \cite{bert,gregor,chitra,ka5}
More importantly, there is an upturn in the susceptibility at $T=75$ K which has been ascribed to the
DM interactions.\cite{rigol,zorko,ofer2}

Many of the above experiments indicate that analyses \cite{cepas,shtengel} of the effect 
of DM interactions on non-magnetic ground states of the Heisenberg Hamiltonian
are needed for a complete understanding of the kagome antiferromagnet. 
For their Dirac algebraic spin liquid, Hermele {\em et al.} \cite{ran2}
showed that the DM coupling was a relevant perturbation, implying that an infinitesimal coupling
induced magnetic order. In a recent exact diagonalization study, Cepas {\em et al.} \cite{cepas} 
reach a different conclusion: they claim that there is a non-zero critical DM coupling $D_c$ beyond
which magnetic order is induced. They estimate $D_c/J \approx 0.1$, quite close
to the value measured\cite{zorko} for ZnCu$_3$(OH)$_6$Cl$_2$ which has $D/J \approx 0.08$.
This proximity led Cepas {\em et al.} to suggest that the quantum criticality of the DM-induced
transition to magnetic order controls the observable properties of this kagome antiferromagent.

\subsection{Hyperkagome lattice}
\label{sec:hyperkagome}

Okamoto {\em et al.} \cite{takagi} have reported that Na$_4$Ir$_3$O$_8$ from a $S=1/2$ 
antiferromagnet on a three-dimensional lattice of corner-sharing triangles, which they called
the hyperkagome. This has no observed magnetic or structural ordering down to the lowest
observed temperatures. Models of a spinon Fermi surface have been proposed \cite{zhou,lawler},
but these overestimate the low $T$ limit of the specific heat $\gamma = C_P/T$; present experiments
do not indicate a significant $\gamma$ in the low $T$ limit. An analysis of a continuous Mott transition
of this spinon Fermi surface state to a Fermi liquid has also been carried out. \cite{podolsky}

\section*{Acknowledgments} 
I thank Reizo Kato and Yasuhiro Shimizu for permission to reproduce Fig.~\ref{fig:kato}, Kirill Shtengel for 
input on Section~\ref{sec:kitaev}, and Ribhu Kaul
for input on Section~\ref{sec:acl}.
This research was supported by the NSF under grant
DMR-0757145, and by the FQXi foundation.

\end{document}